\documentclass{emulateapj-rtx4}
\usepackage{natbib}

\shorttitle{The hard X-ray behavior of Aql~X-1}
\shortauthors{Chen et al.}

\begin{document}

\title{The hard X-ray behavior of Aql~X-1 during type-I bursts}

\author{
Yu-Peng Chen\altaffilmark{1},
Shu Zhang\altaffilmark{1},
Shuang-Nan Zhang\altaffilmark{1},\\
Long Ji\altaffilmark{1},
Diego F. Torres\altaffilmark{2,3},
Peter Kretschmar\altaffilmark{4},
Jian Li\altaffilmark{1},
Jian-Min Wang\altaffilmark{1,5}
}
\email{chenyp@ihep.ac.cn, szhang@ihep.ac.cn}

\altaffiltext{1}
{
Key Laboratory for Particle Astrophysics, Institute of High Energy Physics, Chinese Academy of Sciences, 19B Yuquan Road, Beijing 100049, China
}

\altaffiltext{2}
{
Instituci\'o Catalana de Recerca i Estudis Avan\c cats (ICREA), 08010 Barcelona, Spain
}

\altaffiltext{3}
{
Institute of Space Sciences (IEEC-CSIC), Campus UAB, Torre C5, 2a planta, 08193 Barcelona, Spain
}

\altaffiltext{4}
{
  European Space Astronomy Centre (ESA/ESAC), Science Operations Department,
  Villanueva de la Can{\~n}ada (Madrid), Spain
}


\altaffiltext{5}
{
Theoretical Physics Center for Science Facilities (TPCSF), CAS
}

\begin{abstract}

We report the discovery of an anti-correlation between the soft and the hard
X-ray lightcurves of the X-ray binary Aql~X-1 when bursting.
This behavior may indicate that the corona is cooled by the
soft X-ray shower fed by  the type-I X-ray bursts, and that this process happens
within a few seconds.
Stacking the Aql X-1 lightcurves of type-I bursts, we find a shortage
in the 40--50 keV band, delayed by 4.5$\pm$1.4 s with respect to the soft X-rays.
The photospheric radius expansion (PRE)  bursts are different in that
neither a shortage nor an excess shows up in the hard X-ray lightcurve.

\end{abstract}

\section{Introduction}

A low-Mass X-ray Binary (LMXB) is a system with either a neutron star (NS) or a black hole (BH) accreting material from a companion star (in general having a mass $M \leq M_{\odot}$), usually via a Roche-Lobe overflow  settling into an accretion disk.
Many LMXBs show periods of
  high activities, referred to as outbursts, which are probably triggered by  changes in the mass accretion rate.
   Basically, there are two components (an optically thick blackbody like component in the soft X-ray band and a power law with cutoff component in the hard X-ray band) in the outburst spectrum of a LMXB.
It is thought that the power law component is caused by inverse Comptonization of the hot plasma (i.e. the corona, see the case of 4U~1608$-$522 \citealp{zhang1996}); the seed photons being the
optically thick blackbody like component. However, no consensus has been
reached on the  corona formation model, and both
disk evaporation  \citep{meyer1994,Esin1997,liu2007,frank2002} and magnetic reconnection models \citep{zhang2000,mayer2007,zhang2007} are entertained.
The timescale of the corona formation/heating is key to distinguish between these models.
The typical timescale for disk evaporation and magnetic reconnection models are days \citep{meyer1994,Esin1997,liu2007,frank2002} and milliseconds \citep{zhang2000,mayer2007,zhang2007}, respectively.

Type-I X-ray bursts are caused by unstable burning of the accreted hydrogen/helium on the
surface of a NS enclosed in an XRB, and manifest themselves as a sudden increase (typically by a factor of 10 or greater) in the
X-ray luminosity followed by an exponential decay (for reviews, see \citealp{Lewin,Cumming,Strohmayer,Galloway}).
The most luminous bursts are the photospheric radius expansion (PRE) events,
for which the peak flux is comparable
to the Eddington luminosity at the surface of the NS.

The  spectral behavior of the outbursts (i.e., persistent/accretion emission from the accretion disk and corona)  {might be affected} by the bursts themselves. In fact, a hard X-ray flux decrease was hinted at during a burst of Aquila X-1 (Aql X-1, also named 4U 1908+005) observed with RXTE/HEXTE, but with a significance of only $\sim$ 2 $\sigma$ \citep{maccarone2003}.
 Effects at energies lower than 30 keV were reported by \citet{Worpel2013,Zand2013}.

Aql X-1 is a transient NS XRB, classified as an atoll source \citep{Reig2000}. It is one of the three NS XRBs which show hysteresis in state transition \citep{Gladstone2007}; the other two being
4U 1608-522 \citep{Maitra2004} and IGR J17473-2721 \citep{Zhang2009,chen2010,chen2011}.
Tens of type-I bursts were detected during the outbursts of Aql X-1.
Among them, roughly one third are PRE bursts \citep{Galloway}.
In this paper, we report on a detailed analysis of all RXTE/PCA observations on Aql X-1, with the aim of gathering information about
the timescales of formation of its corona.

\section{Observations and results}

\subsection{Bursts selection}
Aql X-1 was frequently monitored by RXTE during its service till 2012.
 The analysis of the PCA data is performed  using  HEAsoft v.6.6. The data are filtered using the standard RXTE/PCA criteria.
Only the data from the PCU2 (the third Proportional Counters Unit, in the 0-4 numbering scheme) are used for the analysis, because this PCU was 100\% on during all the  observations.
The dead time correction is made to all the spectra and lightcurves following the standard procedure described at the HEASARC
website.

 From all available  RXTE/PCA pointing observations on Aql X-1, we carried out a systematic search for bursts in the lightcurve of Aql~X-1.
Only the bursts observed by PCA in E$\_125$u$\_$64M$\_$1s mode (
 events are time-stamped with 125-microsec resolution, in 64 PHA channel bands with the `M' channel distribution/binning scheme starting at channel 0, and are read out every 1 second)
are chosen for analysis, which have enough energy bands and good time resolution. We picked up 39 bursts to constitute a sample, among which 13  are PRE events (Table 1).

\subsection{Data analysis}

During the bursts, the persistent/accretion and the burst emission are mixed.
In order to investigate  persistent/accretion emission changes
 during the type-I X-ray burst, we analyze  the spectra and lightcurves by  subtracting the
pre-burst emission (including instrumental background and persistent/accretion flux of the neutron star).

The burst spectrum is well modeled by a blackbody
with a characteristic temperature of {a few keV}. The burst emission can reach ${L}_{\rm Edd}$, and
dominates the total emission at energies well below $\sim$40 keV, above which the persistent/accretion emission from the corona takes over.
The average of the 40--50 keV, persistent count rates recorded by RXTE/PCA
for non-PRE and PRE bursts are 0.37$\pm$0.05 cts/s and 0.05$\pm$0.04  cts/s, respectively.


Because the flux in the 40--50 keV band is very faint to be detected  during single bursts,
especially when the  instrument background is $\sim$ 1 cts/s, which is much higher than the persistent/accretion and bursts emission in the same energy band, we stack the individual lightcurves.
 For each burst, we use the time of its peak in the 2--10 keV as a reference to produce the
 lightcurve  of each burst in the 2--10 keV band as well as in the 40--50 keV band.
The  fluxes recorded 48 seconds before
and 80 seconds after the reference time are regarded as the persistent/accretion flux and are subtracted for each burst in timing  analysis.
The results are stable with reasonable changes of the time period where the persistent/accretion count rate is estimated.
After the persistent emission is subtracted,
 we separate the PRE bursts from the non-PRE bursts. 
From the  the Color-color diagram (CCD) and Hardness-Intensity diagram (HID) of Aql X-1 (Fig. \ref{ccd}), the non-PRE bursts and the PRE bursts  are mostly located in  the hard state and the soft state, respectively.
Based on the CCD and HID,
 we subdivide the bursts into non-PRE bursts in the hard state, PRE bursts in  the soft state, non-PRE bursts in the soft state and PRE bursts in the hard state.
 For the latter two groups, the number of  bursts  is too small to draw conclusion (Table 1). In this paper, only the former two groups (the non-PRE bursts in the hard state and the  PRE bursts in the soft state) are  stacked and averaged in each time bin, respectively.

\subsection{Results}

As shown in Fig. \ref{fig1}, for the non-PRE bursts in the hard state, the 40--50
keV flux of the combined 4s-binned lightcurve is mostly negative during the bursts occurrence and flat elsewhere.
 The 40--50 keV decrement  reaches a maximum of -0.9$\pm$0.2
cts/s at the 2-10 keV burst peak, which  amounts to  the whole 40--50 keV persistent flux. A constant fit with a 1 s-bin light curve that ranges from 0 seconds to 15 seconds results in a $\chi^{\rm 2}$ of 74 under 15 dofs, suggesting a significance of 6 $\sigma$ for the shortage.

By assuming that the spectrum of the corona is well described by a cutoff power law (cutoffpow model in XSPEC) with a photon index 1.25, we simulate the PCA  spectra with different cutoff energy,  fixed photon index and  normalization.
We find that the PCA flux in 40--50 keV is 0.37 cts/s, 0.24 cts/s, 0.11 cts/s, 0.06 cts/s and 0.01 cts/s for the cutoff energy  of 40 keV, 30 keV, 20 keV, 15 keV, and 10 keV,  respectively.
This means that the PCA flux between 40 and 50 keV can drop by about one order of magnitude if the corona temperature is cooled from 40 keV to 10 keV, resulting in a significant shortage  at hard X-rays while bursting.
Assuming that the hard X-rays originate from the corona, this
  suggests  that
most of the corona is cooled by the soft photons of the bursts.

 We perform a  cross-correlation
analysis between the two light curves at 2--10 keV and 40--50 keV, with a bin size of  1~s. In this procedure, 1~s-bin lightcurves were adopted  because for a smaller time interval, e.g. 0.5~s, the poor statistics  prevents from estimating a time lag.
The cross-correlation, see Fig. \ref{fig2}, shows that the shortage at 40--50 keV lags that at 2--10 keV by 1.8$\pm$1.5 s. In order to estimate the error, we sampled the two lightcurves by assuming that the flux in each bin has a Gaussian distribution, and estimated the time delay with a cross-correlation method. By sampling the lightcurve a  thousand times, the distribution of the resulted time delay was fitted with a Gaussian to infer the error.

A similar analysis procedure is carried out  on the PRE bursts in the soft state as well. In the  40--50 keV light curves, no hint of a shortage nor an excess is present around the soft X-ray peak. As shown in Fig. \ref{fig1},  the 40--50 keV lightcurve
 hovers around zero during/around the burst. A constant fit to this light curve gives a $\chi^{\rm 2}$ of 9.0 under 5 dofs,
consistent with no deviation.

\section{Discussion}

\subsection{Additional corona cooling by the bursts}

We have found an anti-correlation of the soft and hard X-ray lightcurves of Aql X-1 when bursting, which likely
indicates a cooling of the corona by the soft X-ray showers of the bursts. This reveals a cooling/heating timescale
of less than a few seconds. These results are similar to those previously found for IGR~J17473-2721 \citep{chen2012}, perhaps
hinting at a generic behavior of NS XRBs.

When bursts do not  occur,  the corona cooling is driven by the soft photons from the disk. Therefore a timescale of a few days reflects the actual  timescale of the soft disk photon field, e.g. the  viscous  timescale of the disk.
When bursts happen, their soft photons overwhelm those from the disk and provide additional cooling in a short time (typically of tens of seconds). The anti-correlation between hard and soft X-rays shows that the corona can be cooled and  recovers  quite fast (in seconds).
Such short timescales for recovery is inconsistent with the disk evaporation model in which the formation of a corona is driven and energized by the disk accretion.
Magnetic field reconnection can provide a viable alternative.

\subsection{Persistent/accretion spectrum softens during the bursts}

\begin{figure}[t]
\centering
      \includegraphics[scale=0.2] {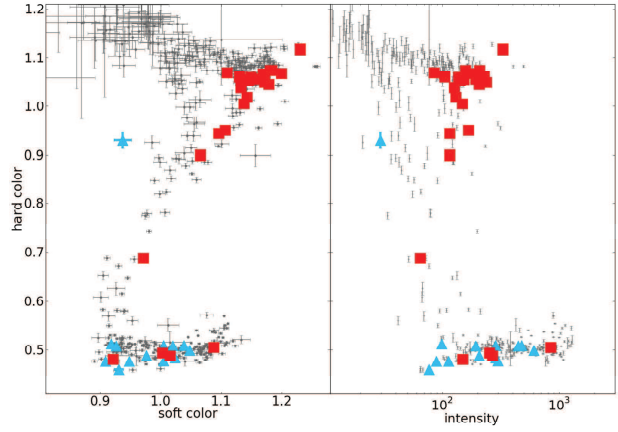}
 \caption{Color-color diagram (the left panel) and Hardness-Intensity diagram (the right panle) for Aql X-1. The soft color is the ratio of the background-subtracted PCA counts in the energy range 3.6-5.0 keV to the counts in the range 2.2-3.6 keV. The hard color is the ratio of counts in the ranges 8.6-18.0 and 5.0-8.6. The intensity is counts in the range 3.6-21.5 keV. The red and blue points represent the position of non-PRE bursts and PRE bursts, respectively. Each point in the diagram corresponds to an average of single OBSID for PCA, which is approximate 3000 seconds.
 }
\label{ccd}
\end{figure}

\begin{figure}[b]
\begin{center}
 \includegraphics[origin=c, angle=0, scale=0.20]{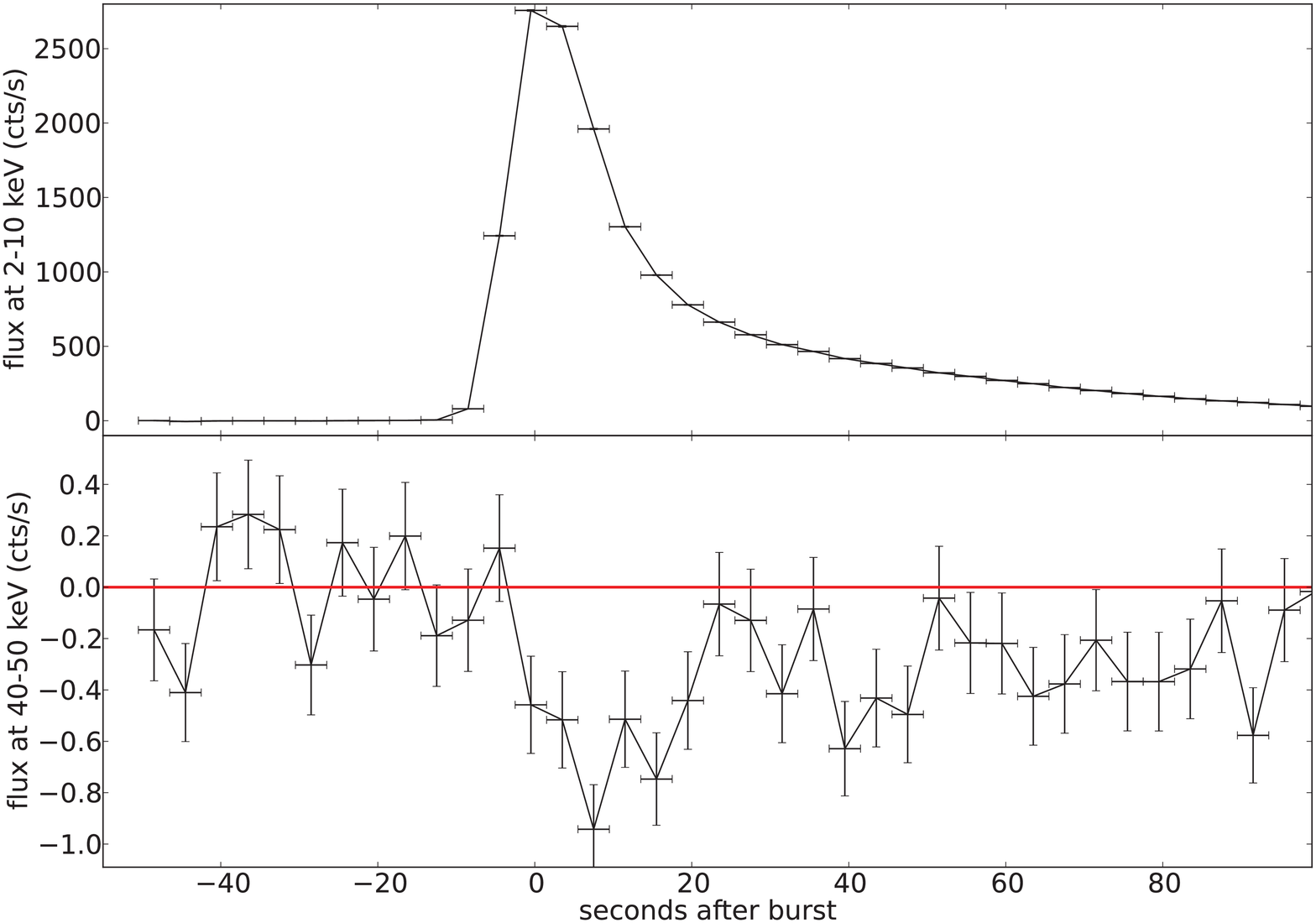}
    \includegraphics[origin=c, angle=0, scale=0.20] {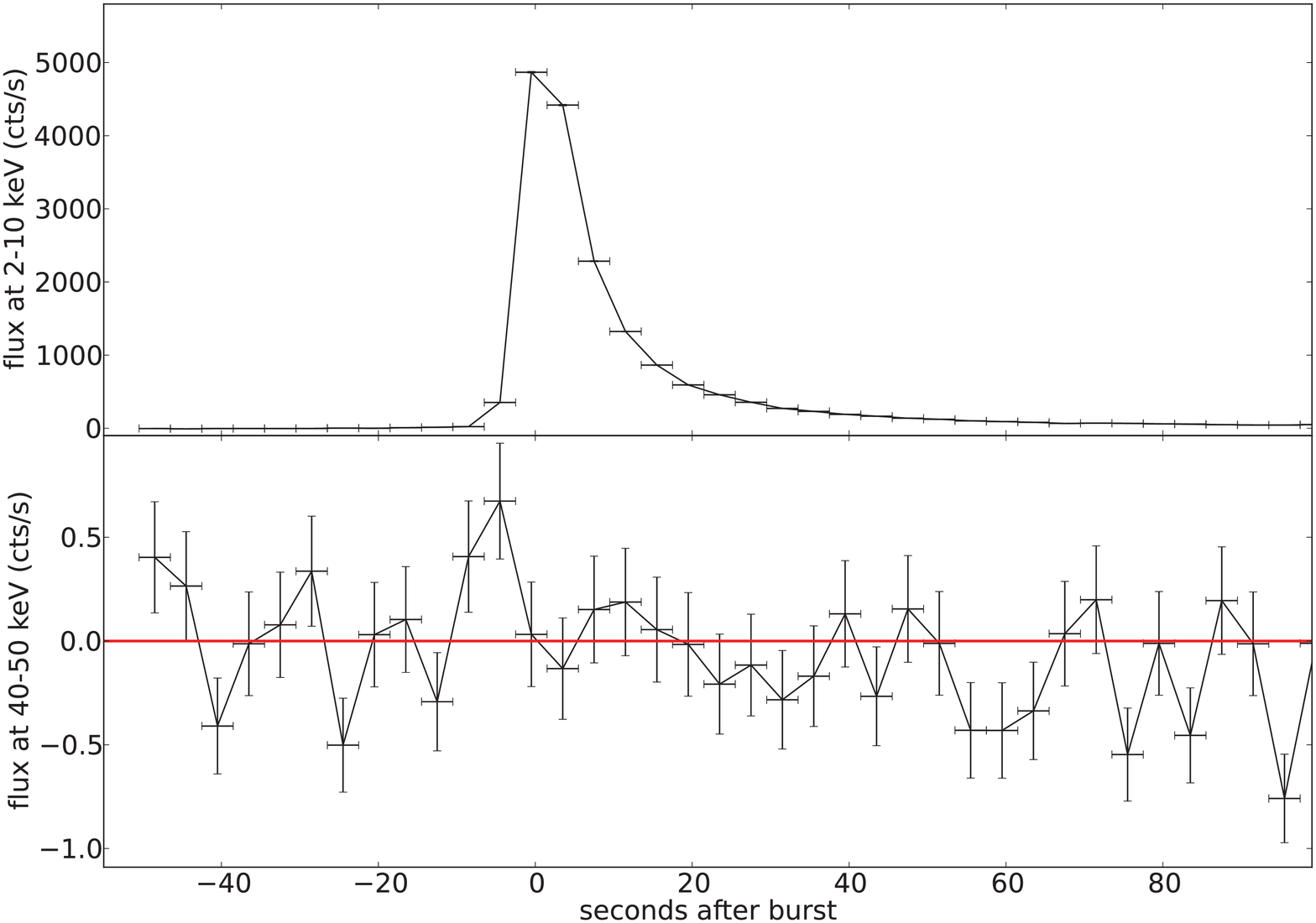}
 \caption{The 4s-binned lightcurves for the non-PRE bursts in the hard state (left two panels) and the PRE bursts in the soft state  (right two panels). Each data point is the sum over the net lightcurve after subtracting off the persistent emissions at 2-10 keV (top panels) and 40-50 keV (bottom panels), respectively.}
   \label{fig1}
\end{center}
\end{figure}

\begin{figure}[t]
\centering
      \includegraphics[angle=270, scale=0.4] {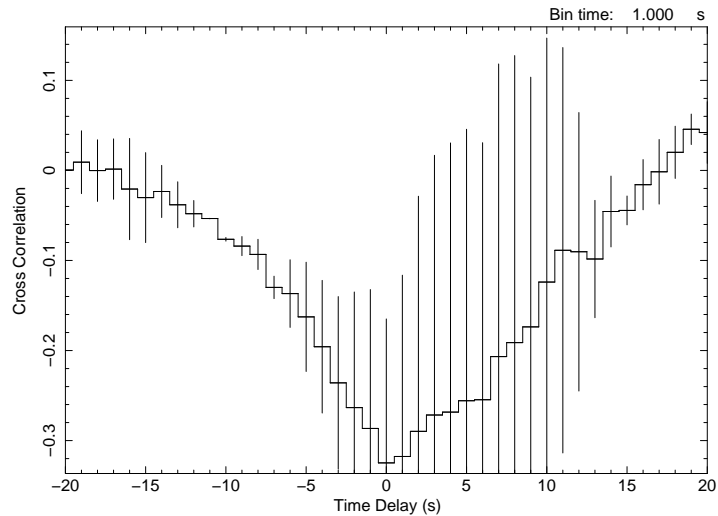}

 \caption{The cross-correlation between the 2-10 keV and 30-50 keV, with a time resolution of 1 second, for the combined non-PRE burst in the hard state.}
\label{fig2}
\end{figure}

The PRE bursts are the most luminous events. Their fluxes can stay very close to the Eddington limit, accompanied with a drop of  color temperature below 1 keV and an increment of the apparent radius  to several tens kilometer.  Although PRE bursts are brighter,
no anti-correlation is found at hard X-rays.
The PRE bursts of Aql X-1 are mostly located in the decaying phase of the outbursts and in the banana state of the CCD diagram (Fig. \ref{ccd}).
The persistent flux at 40--50 keV is quite low (0.05$\pm$0.04 cts/s) during the PRE bursts,
implying   a weak corona and thus poor statistics for measuring a possible
  shortage.

  The persistent/accretion flux in 2.5--25 keV has been found to
 increase by a factor of 20 during the PRE bursts of 40 sources by RXTE/PCA \citep{Worpel2013}.
 In their work, the possible influence of the burst upon the persistent flux was investigated via spectral fitting, where the spectrum was fitted jointly with the burst blackbody and the persistent spectral shape at 2.5--25 keV. A significant increment was derived for the persistent emission during the burst of PRE  events \citep{Worpel2013}, and these authors claim  that this phenomenon is  also detected for the non-PRE bursts in their forthcoming paper.

If the persistent/accretion flux  increased by a factor of 20 during the PRE bursts of Aql X-1, the flux of 40--50 keV ($F_{\rm accretion}$) will be up to 0.05$\times$20 cts/s $\sim$ 1 cts/s. Considering the number of the PRE bursts $N_{\rm PRE}$, the 4-s time bin,  and the
instrument background ($F_{\rm bkg}$$\sim$ 1 cts/s),  the significance of the excess of the each time bin should be
\begin{equation}
\sigma=\frac{F_{\rm accretion}}{{(F_{\rm bkg}+F_{\rm accretion})}^{\rm \frac{1}{2}}}\times (4N_{\rm PRE})^{\rm \frac{1}{2}}=4.5.
\end{equation}
However,  we find no evidence for a hard X-ray increase during the bursts, neither
for IGR~J17473-2721 \citep{chen2012}, nor for
4U~1636-536 \citep{ji2013}, nor for Aql~X-1  (this paper). In contrast, we find a shortage for the non-PRE bursts, which is opposite to the expectation that the accretion rate should increase during bursts.

An enhanced persistent/accretion flux is  also detected in a joint observation by Chandra and RXTE/PCA in 0.5-30 keV \citep{Zand2013}, but with more soft excess  and less hard X-ray emission.
 They suggest that the excess in the spectra during the bursts is due to that the bursts  emission being  reprocessed/reflected by the disk
and re-emitted into the line of sight \citep{Ballantyne2004}.
There  is no conflict between the model above and our finding;
i.e., during  bursts, from the both observations \citep{Worpel2013,Zand2013} in soft X-ray band and our findings in hard X-ray band, there is an increased soft X-ray flux and decreased hard X-ray flux.

\acknowledgements
This work is supported by 973 program 2009CB824800 and the National Natural Science Foundation of China via NSFC-11233003, 11103020, 11133002, 11073021 and 11173023. This work is also done in the framework of the grants AYA2012-39303, SGR2009-811, and iLINK2011-0303.
DFT was additionally supported by a Friedrich Wilhelm Bessel Award of the Alexander von Humboldt Foundation.
This research has made use of data obtained from the High Energy Astrophysics Science
Archive Research Center (HEASARC), provided by NASA's Goddard Space Flight Center.

\bibliographystyle{plainnat}


\begin{table}[ptbptbptb]
\begin{center}
\label{table1}
\caption{
Aql X-1 selected bursts. The columns provide information about the OBSID, time, peak flux in 2-10 keV (in units of cts/s),  on whether
the burst is a PRE one or not and the state when the bursts occurred. }
 \hspace{2pt}
 ¡¡¡¡\renewcommand{\arraystretch}{0.7}
 \small
\begin{tabular}{cccccccccccccccccl}
\hline
 No     & ObsID     &MJD  &F$_{\rm peak}$ (cts/s)    &  PRE  &   $state^{*}$  \\
\hline
 1     & 20098-03-08-00     & 50508.98  & 6983.32    &  yes  &        S        \\
 2     & 20092-01-05-00     & 50696.52  & 7188.68    &  yes  &        S        \\
 3     & 20092-01-05-030       & 50699.40  & 5099.40    &  yes  &        S        \\
 4     & 20092-01-05-07     & 50700.02  & 3776.07    &  no   &        S\\
 5     & 20092-01-05-05     & 50701.54  & 3548.48    &  no   &        S\\
 6     & 40047-03-02-00     & 51332.78  & 7109.68    &  yes  &        S        \\
 7     & 40047-03-06-00     & 51336.59  & 7177.73    &  yes  &        S        \\
 8     & 50049-01-04-02     & 51818.79  & 2963.57    &  no   &        H        \\
 9     & 50049-02-11-00     & 51851.40  & 4964.13    &  no   &        S\\
 10     & 50049-02-13-01     & 51856.16  & 6223.50    &  yes  &        S        \\
 11     & 60054-02-01-01     & 52085.10  & 3419.08    &  no   &        H        \\
 12     & 60054-02-01-02     & 52086.04  & 3296.49    &  no   &        H        \\
 13     & 60054-02-02-01     & 52091.58  & 3227.50    &  no   &        H        \\
 14     & 60054-02-03-03     & 52100.80  & 5658.48    &  yes  &        S        \\
 15     & 60429-01-06-00     & 52324.99  & 6839.94    &  yes  &        S        \\
 16     & 70069-03-02-03     & 52347.18  & 5229.77    &  yes  &        S        \\
 17     & 70069-03-03-07     & 52351.88  & 2701.49    &  no   &        S\\
 18     & 80403-01-05-00     & 53056.12  & 3158.50    &  no   &        H        \\
 19     & 91028-01-07-00     & 53468.29  & 3043.71    &  no   &        H        \\
 20     & 91028-01-09-00     & 53469.13  & 2897.51    &  no   &        H        \\
 21     & 91028-01-12-00     & 53470.99  & 3530.08    &  no   &        H        \\
 22     & 91028-01-13-00     & 53471.76  & 2795.50    &  no   &        H        \\
 23     & 91028-01-14-00     & 53472.21  & 2781.69    &  no   &        H        \\
 24     & 91028-01-18-00     & 53474.51  & 2525.70    &  no   &        H        \\
 25     & 91028-01-20-00     & 53476.03  & 2624.96    &  no   &        H        \\
 26     & 91028-01-21-00     & 53477.00  & 2761.38    &  no   &        H        \\
 27     & 91028-01-21-00        & 53477.01  &  992.11    &  no   &        H        \\
 28     & 91414-01-08-00     & 53715.16  & 3141.22    &  no   &        H        \\
 29     & 91414-01-08-03     & 53719.43  & 3312.39    &  no   &        H        \\
 30     & 91414-01-09-00     & 53720.16  & 3387.41    &  no   &        H        \\
 31     & 93076-01-03-00     & 54245.56  & 2395.18    &  no   &        H        \\
 32     & 93076-01-09-00     & 54251.53  & 3979.26    &  no   &        H        \\
 33     & 92438-01-02-01     & 54259.25  & 8079.84    &  yes  &        T\\
 34     & 93405-01-03-07     & 54365.81  & 8005.16    &  yes  &        S        \\
 35     & 94076-01-04-01     & 55149.08  & 2966.77    &  no   &        H        \\
 36     & 94076-01-04-03     & 55151.19  & 3321.13    &  no   &        H        \\
 37     & 94076-01-05-02     & 55157.14  & 6655.43    &  yes  &        S        \\
 38     & 96440-01-09-07     & 55904.23  & 5830.69    &  yes  &        S        \\
 39     & 96440-01-09-01     & 55905.33  & 3577.68    &  no   &        T        \\
\hline
\hline
\end{tabular}
\end{center}
\begin{list}{}{}
\item[]{State$^{*}$ when the bursts occurred, S=soft state, H=hard state, T= transition state.}
\end{list}
\end{table}



\end{document}